\documentclass[pdflatex,sn-mathphys-num]{sn-jnl}
\usepackage{tabularx}
\usepackage{rotating}
\usepackage{graphicx}%
\usepackage{multirow}%
\usepackage{amsmath,amssymb,amsfonts}%
\usepackage{amsthm}%
\usepackage{mathrsfs}%
\usepackage[title]{appendix}%
\usepackage{xcolor}%
\usepackage{textcomp}%
\usepackage{manyfoot}%
\usepackage{booktabs}%
\usepackage{algorithm}%
\usepackage{algorithmicx}%
\usepackage{algpseudocode}%
\usepackage{listings}%
\usepackage{graphicx}
\usepackage{booktabs} 
\usepackage{caption}  
\usepackage{graphicx}
\usepackage{amsmath,amssymb,amsfonts}
\usepackage{booktabs}

\usepackage{siunitx} 

\theoremstyle{thmstyleone}%
\theoremstyle{thmstyletwo}%

\theoremstyle{thmstylethree}%

\begin{document}

\title[Article Title]{Prediction of Cellular Malignancy Using Electrical Impedance Signatures and Supervised Machine Learning }


\author*[1]{\fnm{Shadeeb} \sur{Hossain}}\email{shadeeb@shadeebengineeringlab.com}


\affil*[1]{\orgdiv{Research Division}, \orgname{Shadeeb Engineering Lab}, \orgaddress{\street{New York}, \city{Brooklyn}, \postcode{11223}, \state{NY}, \country{USA}}}




\abstract{Bioelectrical properties of cells such as relative permittivity, conductivity, and characteristic time constants vary significantly between healthy and malignant cells across different frequencies. These distinctions provide a promising foundation for diagnostic and classification applications. This study systematically reviewed 20 scholarly articles to compile 535 datasets of quantitative bioelectric parameters in the kHz-MHz frequency range and evaluated their utility in predictive modeling. Three supervised machine learning algorithms- Random Forest (RF), Support Vector Machine (SVM), and K-Nearest Neighbor (KNN) were implemented and tuned using key hyperparameters to assess classification performance.  In the second stage, a physics informed framework was incorporated to derive additional dielectric descriptors such as imaginary permittivity, loss tangent and charge relaxation time from the measured parameters. Random Forest based feature importance analysis was employed to identify the most discriminative dielectric parameters influencing the classification process. The results indicate that dielectric loss related parameters, particularly imaginary permittivity and conductivity, contribute significantly to the classification of cellular states. While the incorporation of physics-derived features improves model interpretability and reduces overfitting tendencies, the overall classification accuracy remains comparable to models trained using primary dielectric descriptors. The proposed approach highlights the potential of physics-informed machine learning for improving the analysis of dielectric spectroscopy data in the biomedical diagnostics. 
}




\maketitle

\section{Introduction}\label{sec1}

According to the Centers for Disease Control and Prevention (CDC) and the National Cancer Institute, approximately 614,000 cancer-related deaths were reported in the United States in 2023. Additionally, an estimated 1.9 million and 2.0 million new cancer cases were reported in 2022 and 2025, respectively. Several factors influence cancer prognosis including cancer type, (ii) stage at diagnosis, (iii) cellular characteristics, and (iv) histological grade, among others \cite{Rakha2010}. Notably, changes in the electrical properties and biochemical composition of cells have emerged as significant indicators for disease diagnosis and progression \cite {Hossain2020}. Early detection is strongly correlated with improved survival outcomes, underscoring the importance of timely diagnosis and monitoring to reduce mortality and expand treatment options \cite{McPhail2015},\cite{Muller2018},\cite{vandenBergh2015}. 

One of the key advantages of electrical characterization of cells is its non-invasive and label-free nature \cite{NgocLe2019}. Unlike fluorescence -based techniques, dielectric measurement does not require staining or labeling, which can potentially alter cellular behavior. Moreover, cells remain viable after measurement, enabling further downstream analysis or long-term studies if needed. Another notable advantage, driven by advancements in medical technology, is that real-time dynamic electrical measurements enable continuous cell monitoring and facilitate high-throughput screening. 

Cheung et al. (2005) developed a microfabricated impedance spectroscopy flow cytometer for the rapid dielectric characterization of cells across a range of frequencies \cite{Cheung2005}. They utilized electrical features such as amplitude, phase, and opacity to actively discriminate between cell types without the use of molecular markers. Although the study was limited to red blood cells (RBC), the approach holds potential for broader applications, including the measurements of cell conductivity and capacitance for early-stage apoptosis or cancer diagnostics.  

Similarly, Gawad et al. (2001) introduced a cytological tool capable of high-speed cell counting and separation, operating at a sampling rate of 100 samples per second \cite{Gawad2001}. Their chip-based flow cytometer employed a differential pair of microelectrodes to measure impedance over a frequency range of 100 kHz to 15 MHz. This system can be extended to distinguish between normal and malignant cells based on differences in bioimpedance or conductivity. 

Our previous works focused on electrical impedance and transmembrane potential as biophysical indicators for cancer cell diagnosis \cite{Hossain2021},\cite{Hossain2020}.  The study revealed that cancer cells exhibit higher electrical conductivity and greater dielectric loss compared to their healthier counterparts. Additionally, cancer cells tend to be depolarized, often displaying elevated intracellular sodium ion concentrations despite similar external ionic conditions \cite{Kunzelmann2005},\cite{Bortner2014},\cite{Binggeli1986}. For instance, ovarian tumor cells have reported to possess a transmembrane potential of approximately -5 mV, while MCF -7 breast cancer cells range between -37 to -38 mV , and MDA- MB-468 cells exhibit a potential around -30 mV \cite{Mahmoud2023}.
In our previous works, we also investigated that both the bioimpedance and optical properties of breast cancer cells can be used to enhance the precision of malignant cell identification \cite{Hossain2020},\cite{Hossain2021},\cite{HossainHossain2021}. Electrical measurements were conducted in the microwave frequency range, while the optical characteristics of normal and cancer cells were evaluated in the visible and near- infrared spectral (NIR) regions.  Similarly, Alfano et al. conducted several pioneering studies utilizing optical techniques to distinguish precancerous and cancerous tissues based on their characteristic absorption wavelength \cite{Alfano2016},\cite{Alfano2013},\cite{Alfano2005}.  However, despite their potential, optical diagnostic techniques face significant challenges, including :(i) limited spatial resolution, (ii) prolonged acquisition time, and (iii) reduced sensitivity in complex biological environments \cite{Licha2005}. 
While numerous studies have demonstrated the feasibility of utilizing raw impedance spectra or optical signatures for cancer detection, most existing approaches rely primarily on directly extracted measurement features, which often lack clear physical interpretability and generalizability across experimental platforms. To address this limitation, there is a growing need for frameworks that integrate electrical measurements with biophysical models to derive intrinsic cellular properties-such as membrane permittivity, cytoplasmic conductivity and transmembrane potential that more directly reflect underlying cellular physiology. Such physics guided representation can provide robust, interpretable biomarkers and form a stronger foundation for data driven learning algorithm. 

 Measurable bioelectrical differences between cancerous and non-cancerous cells provide a valuable feature space for computational classification. When combined with equivalent circuit model and electromagnetic analysis, these measurements can be transformed into physically meaningful dielectric and conductive parameter.By leveraging machine learning algorithms, such as Support Vector Machine (SVM), Random Forest (RF), Deep Neural Network (DNN) or K-Nearest Neighbor (KNN), it is possible to learn patterns in electrical impedance, dielectric loss and transmembrane potential that distinguish malignant cells from healthy ones \cite{Biau2016},\cite{Rigatti2017},\cite{Guo2003},\cite{Hossain2024Authorea},\cite{Suthaharan2016}. These models can be trained on labeled datasets derived from impedance cytometry or microelectrode array recordings, enabling real time non-invasive cancer diagnostics with high accuracy. Moreover, the ability of machine learning to capture complex, non-linear relationships enhance the diagnostic potential beyond traditional threshold-based methods.  
 
 In this work, a comparative investigation is conducted to evaluate the performance of these machine learning (ML) algorithms for the classification of biological cell types using dielectric spectroscopy data measured across the kHz-MHz frequency range. In the first stage, the ML models are trained using primary dielectric properties; namely relative permittivity and electrical conductivity to assess their baseline classification capability based on experimentally measured parameters. 
 
In the second stage, a physics-informed framework is incorporated to derive additional dielectric descriptors including: the charge relaxation time ($\tau$), loss tangent ($\tan \delta$), and imaginary permittivity ($\varepsilon''$). These parameters are associated with dielectric relaxation phenomena and energy dissipation mechanisms within biological tissues. The influence of these physics-derived features on the classification performance of the ML models is systematically investigated under identical hyperparameter configuration. Furthermore, a feature importance analysis based on the Random Forest algorithm is performed to identify the most discriminative dielectric parameters contributing to accurate cell classification. The proposed analysis therefore provides insight into the role of physics-informed dielectric parameters in improving machine-learning-based characterization of biological cells.

\begin{figure}[ht]
    \centering
    \includegraphics[width=1.2 \textwidth]{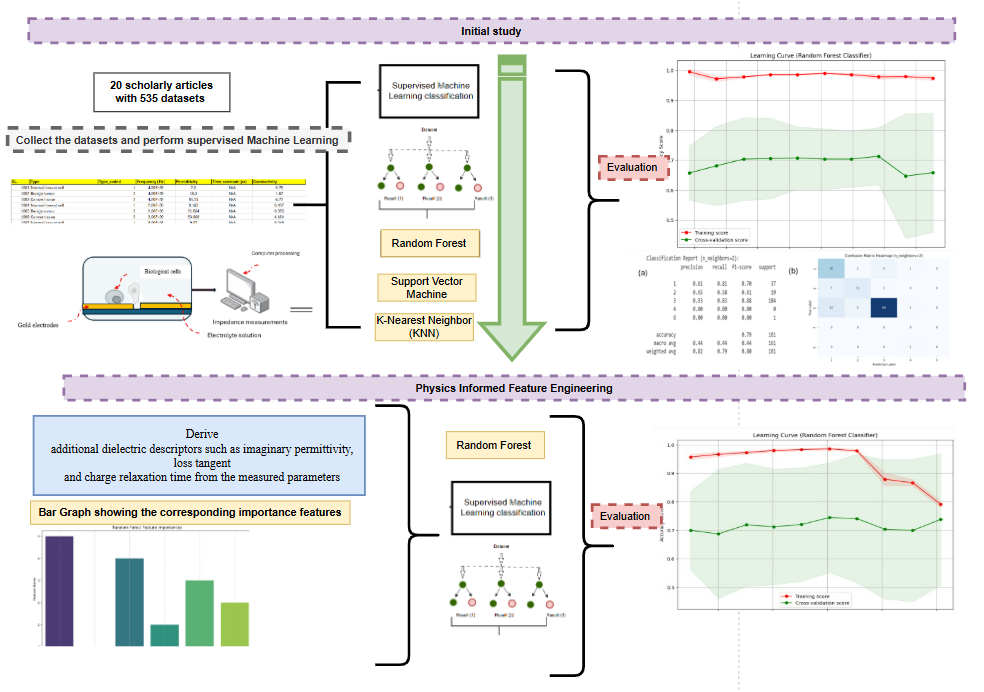}
    \caption{Overall workflow of the proposed methodology for dielectric spectroscopy-based cell classification. (a) Initial machine learning study using experimentally reported dielectric parameters compiled from multiple studies, including frequency, relative permittivity, and electrical conductivity. Supervised learning algorithms such as Random Forest, Support Vector machine , and K-Nearest Neighbor are used for classification and evaluation. (b) Physics informed feature engineering framework where additional dielectric descriptors such as imaginary permittivity, loss tangent and charge relaxation time are derived from the measured parameters. Feature importance analysis is then applied to identify the influential descriptors before retaining the machine learning model for improved classification. }
    \label{fig:biomarkers31}
\end{figure}

The overall workflow of the proposed physics-informed framework is illustrated in Fig.~\ref{fig:biomarkers31}. 

This study investigated and compared the performance of three supervised machine learning models- Random Forest (RF), Support Vector Machine (SVM) and K-Nearest Neighbor (KNN)- for classifying cell samples based on their bioimpedance properties. Labelled datasets were compiled from 20 scholarly journals with 535 datasets, and model hyperparameters were systematically tuned to enhance predictive accuracy. Performance was evaluated using accuracy and F1 score, with the model achieving the highest metric proposed as a potential reference standard for malignancy identification. 

This paper is structured as follows: (i) Background Information - introduces the Cole-Cole plot and provides overview of the three supervised machine learning models; (ii) Data Collection and Methodology – discusses about the source of data and approach (iii) Results and Discussion- compares the models’s evaluation metrics and examines the influence of hyperparameter tuning; (iv) Conclusion and Future Work- compares the model’s evaluation metrics and examines the influence of hyperparameter tuning. 

\section{Background Information}\label{sec2}
\subsection{Bioelectrical Properties and Cole-Cole plot}\label{subsec2}
Cancer progression is accompanied by distinct biophysical and biochemical changes at the cellular level, many of which affect the structural and electrical properties of the cell membrane and cytoplasm \cite{Hossain2020},\cite{Hossain2021}. As a result electrical characterization techniques have emerged as a valuable tool in cancer research, offering non-invasive label-free methods to distinguish between normal, premalignant and malignant cells. Among these techniques, dielectrophoresis (DEP) has shown promise in detecting subtle differences in detecting dielectric properties such as permittivity, conductivity and membrane capacitance \cite{Pethig2010},\cite{Pohl1971},\cite{Nakano2013},\cite{Hossain2023},\cite{Hossain2020},\cite{Lapizco2007}. These properties which are frequency dependent can be modeled using frameworks like the Cole-Cole equation to reveal patterns that correlate with malignancy. The ability to extract quantitative dielectric signatures makes the approaches particularly suitable for early detection, classification, and monitoring of cancer progression in vitro and in vivo settings.  

\begin{figure}[ht]
    \centering
    \includegraphics[width=1.2\textwidth]{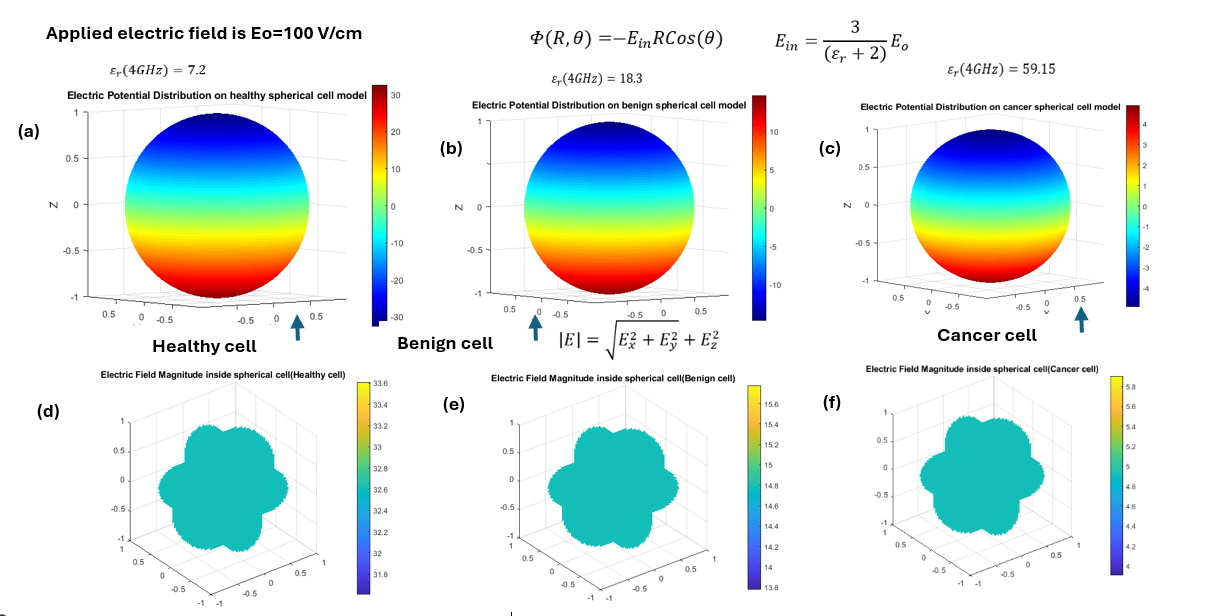}
    \caption{Quantitative simulation models comparing three homogeneous baseline cell representations: (a) healthy, (b) benign, and (c) cancerous, illustrating the influence of bulk dielectric properties on the internal electric field. The corresponding electric field magnitude distributions inside the dielectric sphere are shown for (d) healthy, (e) benign, and (iii) cancerous homogeneous cell.}
    \label{fig:biomarkers}
\end{figure}

Fig.~\ref{fig:biomarkers} (a-f) compares three homogeneous baseline spherical cell models representing healthy, benign, and cancerous cell models to examine the influence of bulk dielectric properties on the internal electric field response. Fig.~\ref{fig:biomarkers} (a-c) shows the quantitative distribution of the scalar electric potential on the surface of the dielectric sphere for each cell type. The surface potential depends on: (i) the relative permittivity of the bulk cell, (ii) the magnitude of the externally applied electric field, and (iii) angular orientation with respect to the applied field direction.

Mathematically, the internal electric field ($E_{in}$) and surface potential ($\Phi$) of a homogeneous dielectric sphere subjected to a uniform external electric field are given by equations \eqref{eq:internal_field} and \eqref{eq:potential}:
\cite{Ref65},\cite{Ref66}:
\begin{equation}
    E_{in} = \left( \frac{3}{\epsilon_r + 2} \right) E_o
    \label{eq:internal_field}
\end{equation}

\begin{equation}
    \Phi(r, \theta) = -E_{in} R \cos \theta
    \label{eq:potential}
\end{equation}

where $E_{in}$ is the uniform internal electric field inside the dielectric sphere, $E_o$ is the applied external field, $R$ is the radius of the cell, $\epsilon_r$ is the relative permittivity of the bulk medium, and $\theta$ is the polar angle measured from the field direction. 

Relative permittivity values of 7.2, 18.3 and 59.15 were used to represent healthy, benign and cancerous cells, respectively. As the relative permittivity increases, a reduction in the surface electric potential magnitude is observed, indicating enhanced dielectric screening with the cell.

Fig. ~\ref{fig:biomarkers} (d--f) illustrates the corresponding electric field magnitude distribution inside the homogeneous spheres calculated using equation \eqref{eq:magnitude}:

\begin{equation}
    |E| = \sqrt{E_x^2 + E_y^2 + E_z^2}
    \label{eq:magnitude}
\end{equation}

where $E_x$, $E_y$, and $E_z$ are the Cartesian components of the electric field.

The electric potential exhibits a dipolar distribution characteristic of a polarized dielectric sphere, while the electric field magnitude remains spatially uniform within each homogeneous model. This behavior is consistent with the analytical solution of Laplace’s equation for a dielectric sphere embedded in a uniform electric field. The homogeneous model isolates bulk permittivity-driven field screening effects, providing a useful baseline for comparison with impedance-based measurements relevant to cellular diagnosis. However, real biological cells possess a thin insulating membrane and exhibit frequency-dependent conductivity and interfacial polarization effects, which are not captured in this simplified representation.

A biological cell can be electrically modeled as a combination of resistive and capacitive elements \cite{Brosseau2021},\cite{Gowrishankar2003},\cite{Ellappan2005}. The cytoplasm, rich in ions, is represented by an internal resistance, Ri reflecting its conductive properties. In contrast, the cell membrane, primarily composed of a lipid bilayer, acts as a dielectric barrier and is modeled as a parallel combination of resistance and capacitance. This RC configuration accounts for the membrane’s ability to store and partially block charge flow. Importantly, the overall importance of the cell is frequency-dependent: at low frequencies, the capacitive nature of the membrane dominates, impending current flow; while at higher frequencies, the capacitive reactance decreases, allowing current to pass through resistive pathways more easily \cite{Trainito2015},\cite{Mahesh2020},\cite{Karmakar2024}.  

 Iqbal et al. (2019) presented an equivalent electrical model of an isolated single biological cell composed of resistive and capacitive elements \cite{Iqbal2019}. In this model, the cell is represented by distinct compartments corresponding to the cytoplasm, nucleus and cell membrane. The capacitive components arise primarily from the charge-separation layers such as cell membrane and nuclear envelope, while the cytoplasm and nucleus are modeled as lossy conductive media and therefore contribute predominantly resistive components. 
 
Electrical impedance measurements of biological cells reflect the combined contribution of resistive and capacitive elements. Effective capacitance is governed by the relative permittivity of cellular constituents, which is inherently frequency-dependent and influences the measured impedance response. Cancerous cells are reported to exhibit higher relative permittivity compared to healthy cells, leading to enhanced capacitive effects. These dielectric contrasts form the physical basis for impedance-based differentiation of cellular states. 

\begin{figure}[ht]
    \centering
    \includegraphics[width=1.0\textwidth]{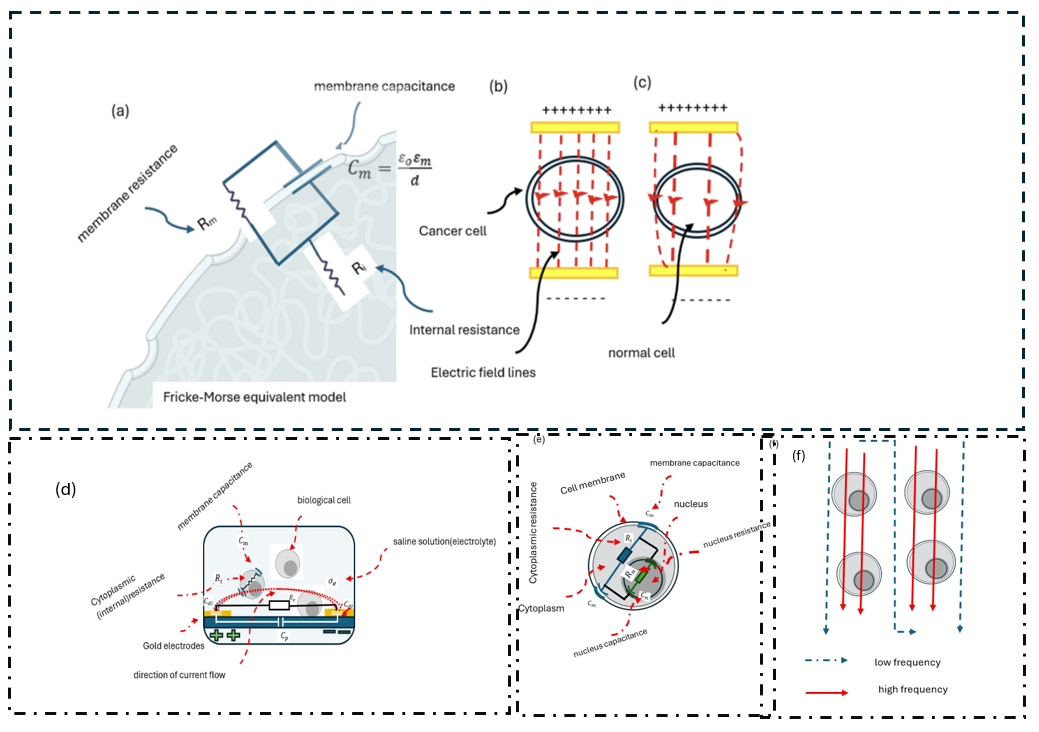} 
   \caption{\small{(a)}: Illustration of the Fricke-Morse equivalent circuit model representing a biological cell. (b) and (c): Variation in electric field distribution highlighting difference between normal cells and cells with higher effective permittivity (e.g. cancerous cells). (d) Schematic of impedance measurements using planar gold electrodes immersed in a conductive electrolyte medium, (e) Extended equivalent circuit model incorporating resistive and capacitive elements representing the membrane, cytoplasm and nucleus. (f) Frequency-dependent electrical behavior of a cell, illustrating current confinement at lower frequencies and membrane penetration at higher frequencies. }
    \label{fig:biomarkers11}
\end{figure}

Giannoukis and Min (2014) explored the mathematical and physical modeling of the dynamic electrical bioimpedance of cells \cite{Giannoukos2014}. Fig. ~\ref{fig:biomarkers11} (a) illustrates the Frickle-Morse equivalent circuit model representing a biological cell. Fig.~\ref{fig:biomarkers11} (b) and (c) demonstrate the variation in electric field distribution within normal and cancerous cells, representing how differences in dielectric properties influence the field patterns. The increased permittivity of cancerous cells alters the dielectric response, resulting in more concentrated electric field distribution compared to non-tumorous.

Fig. 3(d) illustrates the schematic of an impedance-based cell analysis technique, in which a small AC excitation current is applied across microelectrodes immersed in a conductive saline medium containing suspended cells. The presence of suspended cells perturbs the resulting current flow due to contrasts in their frequency-dependent dielectric properties, enabling the extraction of electrical impedance in a label free and real-time manner. This approach has been widely employed for dynamic cell analysis and characterization \cite{Iqbal2019},\cite{Ref68},\cite{Ref69},\cite{Ref70}. Under the assumption of a dilute cell suspension and a simplified single-shell electrical representation, the effective impedance of the cell-electrolyte system can be approximated using equation \eqref{eq:5}\cite{Ref68}.

   \begin{equation}
    Z_{mix} = \frac{R_e (1 + j\omega R_i C_m)}{j\omega R_e C_m + (1 + j\omega R_i C_m)(1 + j\omega R_e C_m)}
    \label{eq:5}
\end{equation}
where $R_e$ is the resistance of the electrolyte solution, $\omega$ is the angular frequency, $R_i$ is the cytoplasmic resistance, and $C_m$ is the membrane capacitance.

The frequency-depending dielectric contrast between the cell membrane and intracellular component leads to a measurable variation in current flow, from which the impedance spectrum can be extracted \cite{Ref70}. In Electric cell-substrate impedance sensing (ECIS), impedance is defined as the ratio of the measured voltage  to the applied current across the electrodes over a range of AC frequencies\cite{Ref71}. 
Fig. 3 (e) presents a complex equivalent circuit model of a biological cell inspired by the Fricke-Morse single-shell model. The cytoplasm, owing to its intracellular conductivity is represented by an intracellular resistance $R_i$ while  both the cell membrane and nucleus are modeled using parallel resistive and capacitive elements, characterized by membrane capacitance $C_m$ and nuclear capacitance $C_n$ , respectively. These capacitive components govern the frequency-dependent electrical behavior of the cell through capacitive coupling, as illustrated in Fig. 3(f). 
The capacitive reactance of the cell membrane is given by equation \eqref{eq:6}:

\begin{equation}
    X_c = \frac{1}{2\pi f C_m}
    \label{eq:6}
\end{equation}

At higher frequencies, the membrane reactance $X_{\text{c}}$ decreases, allowing electric current to penetrate the cell interior, whereas at lower frequencies the increased reactance inhibits the current flow through the membrane, confining it primarily to the extracellular region. 

Several impedance-based techniques have been developed to protect the electrical properties of biological cells, each differing in electrode configuration, frequency range, sensitivity, and application domain. These methods exploit the frequency-dependent dielectric response of cell membranes and intracellular components to extract distinctive electrical impedance signatures associated with cellular morphology and physiological state. Table \ref{tab:model_comparison_landscape}summarizes representative impedance measurements techniques reported in the literature, highlighting the cell types studied, key electrical parameters extracted, and their reported applications in tumor cell characterization.

\begin{sidewaystable}[ph] 
\centering
\caption{Comparison of impedance-based measurement techniques and reported electrical parameters of tumor cells from the literature}
\label{tab:model_comparison_landscape}
\renewcommand{\arraystretch}{1.5} 
\small 
\begin{tabularx}{\linewidth}{@{} l >{\raggedright\arraybackslash}X >{\raggedright\arraybackslash}X >{\raggedright\arraybackslash}X l >{\raggedright\arraybackslash}X l @{}}
\toprule
\textbf{Technique} & \textbf{Method} & \textbf{Application} & \textbf{Study} & \textbf{Freq.} & \textbf{Results} & \textbf{Ref.} \\ \midrule

Electric Cell Impedance Sensing (ECIS) & 
Measures cell-electrode impedance by monitoring voltage response to a known applied AC current. & 
Monitoring cell adhesion, migration, and cytotoxicity \cite{Ref72}. & 
OVCA 429 ovarian cancer cells & 
10 Hz--100 kHz & 
$R_b$: 152$\pm$59 $\Omega\cdot$cm$^2$; $C_m$: 8.5$\pm$2.4 $\mu$F/cm$^2$ & 
\cite{Ref73} \\ \addlinespace

Impedance Flow Cytometer (IFC) & 
Interdigitated microelectrodes measure impedance changes of single cells flowing through a channel. & 
Single-cell electrical characterization and biomolecule detection. & 
HepG2, A549, and HeLa cells & 
1 MHz & 
HepG2: 44.6$\pm$10.9 k$\Omega$; A549: 34.9$\pm$12.6 k$\Omega$; HeLa: 26.7$\pm$11.7 k$\Omega$ & 
\cite{Ref74} \\ \addlinespace

Electrical Impedance Spectroscopy (EIS) & 
Analyzes complex impedance spectrum using equivalent circuit models over a broad range. & 
Label-free characterization of dielectric properties and cell discrimination. & 
SW403, Jurkat, and THP-1 & 
10 Hz--1 MHz & 
$R_{ct}$ (SW403/Jurkat): 2250 $\Omega$; THP-1: 2000 $\Omega$ & 
\cite{Ref75} \\ \addlinespace

Microfluidic cell impedance analysis & 
Measures voltage/impedance changes as single cells flow through microchannels under AC excitation. & 
Label-free discrimination based on size, membrane, and conductivity. & 
A549 tumor cells, RBC, and WBC & 
0.1--10 MHz & 
A549: 4.2--79.9 mV; RBC: 2.4--19 mV; WBC: 1.7--10.5 mV & 
\cite{Ref76}\\ \bottomrule

\end{tabularx}
\end{sidewaystable}

Table\ref{tab:cell_parameters} compares electrical impedance parameters, including resistive and capacitive elements, reported for healthy and cancerous cells across different impedance-based measurement techniques and frequency ranges. These parameters play a critical role in characterizing cellular electrical behavior for diagnostic applications. A consistent trend observed across multiple studies is that healthy cells generally exhibit higher membrane capacitance compared to cancerous cells, with capacitance decreasing progressively as cellular invasiveness increases. This reduction is commonly attributed to alterations in membrane morphology, composition, and surface roughness associated with malignant transformation. Similarly, the magnitude of cellular resistance decreases in cancerous cells, reflecting increased ionic conductivity and membrane permeability resulting from disputed cellular architecture. Together, these impedance variations provide discriminative electrical signatures that can be exploited for cancer detection and phenotypic classification.

The frequency-dependent dielectric behavior of biological tissues and cells is commonly described using the complex permittivity formalism:
\begin{equation}
\epsilon^*(\omega) = \epsilon'(\omega) - j\epsilon''(\omega)
\label{eq:complex_permittivity}
\end{equation}
where $\epsilon'$ represents the real (energy storage) component and $\epsilon''$ denotes the imaginary (loss) component of permittivity.

To model dielectric relaxation in heterogeneous biological media, the Cole-Cole dispersion model is employed:
\begin{equation}
\epsilon^*(\omega) = \epsilon_\infty + \frac{\epsilon_s - \epsilon_\infty}{1 + (j\omega\tau)^{1-\alpha}} + \frac{\sigma_i}{j\omega\epsilon_0}
\label{eq:cole_cole}
\end{equation}
where $\epsilon_s$ and $\epsilon_\infty$ denote the static and high-frequency permittivity limits, respectively, $\tau$ is the characteristic relaxation time, $\alpha$ ($0 \leq \alpha \leq 1$) describes the distribution of relaxation times, $\sigma_i$ represents the ionic conductivity, and $\epsilon_0$ is the permittivity of free space.

For $\alpha = 0$, the model reduces to classical Debye relaxation. Nonzero $\alpha$ values account for dispersion broadening associated with structural heterogeneity, membrane polarization, and intracellular complexity. In biological tissues, variations in water content, membrane integrity, and ionic composition significantly influence these parameters, leading to distinguishable trajectories in the complex permittivity plane~\cite{Ref85, Ref86}.

\begin{figure}[ht]
    \centering
    \includegraphics[width=1.1\textwidth]{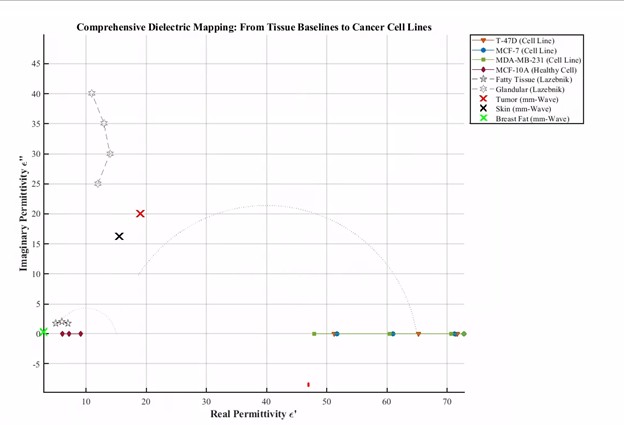} 
    \caption{\small{Cole-Cole complex permittivity mapping of malignant and non-tumorigenic breast cell lines.}} 
    \label{fig:biomarkers2}
\end{figure}

The Cole-Cole representation therefore provides a physically grounded framework for comparing dielectric signatures for malignant and non-tumorigenic breast cell lines, enabling visualization of dispersion characteristics and subsequent quantitative classification analysis. The Cole-Cole relaxation model provides a robust framework for characterizing dielectric dispersion in biological tissues~\cite{Cole1941}. Fig.~ \ref{fig:biomarkers2} illustrates the complex permittivity distribution of malignant (T-47D, MCF-7, MDA-MB-231) and non-tumorigenic MCF-10A breast epithelial cell lines, alongside reported tissue baselines~\cite{Ref82, Ref83, Ref84}. The theoretical Cole-Cole relaxation loci (gray arcs) are superimposed to conceptualize the dispersion behavior.

A distinct clustering pattern is observed in the complex permittivity plane in Fig.~\ref{fig:biomarkers2}: low-water content fatty tissues occupy the lower dielectric regimes, whereas malignant cell lines consistently exhibit elevated real permittivity values ($\epsilon'$) across the investigated frequency range. Importantly, the tumor-fat dielectric contrast persists into the millimeter-wave regime, supporting the broadband viability of electromagnetic diagnostic techniques. 

The spatial distribution of the dielectric signatures visually suggests potential separability in the complex permittivity domain. Quantitative classifier analysis (e.g., linear discrimination analysis and Support Vector Machines) is performed in the following section to rigorously evaluate discriminative performance.

\begin{sidewaystable}[p]
\centering
\caption{Summary of reported resistive and capacitive electrical parameters of healthy and cancerous cell types from the literature.}
\label{tab:cell_parameters}
\renewcommand{\arraystretch}{1.3} 
\small
\begin{tabularx}{\linewidth}{@{} l l l >{\raggedright\arraybackslash}X l l @{}}
\toprule
\textbf{Cell Type} & \textbf{Nature} & \textbf{Frequency} & \textbf{Electrical Parameter} & \textbf{Value/Units} & \textbf{Ref.} \\ \midrule

HT29 + 10\% SW48 & Cancerous & 20--200 kHz & Impedance & 1100--200 $\Omega$ & \cite{Ref77} \\
HT29 + 10\% SW48 & Cancerous & 20--200 kHz & Impedance & 1460--300 $\Omega$ & \cite{Ref77} \\
SW 48 & Cancerous & 20--200 kHz & Impedance & 45--5 k$\Omega$ & \cite{Ref77} \\ \addlinespace

MDCK II cells & Healthy & 500 Hz--64 kHz & Resistance & 5--105 k$\Omega$ & \cite{Ref78} \\
MDCK II cells & Healthy & 500 Hz--64 kHz & Capacitance & 3.9--0.66 nF & \cite{Ref78} \\ \addlinespace

MDA-MB-231 & Cancerous (High) & 20--101 kHz & Resistance & 6.82--3.66 k$\Omega$ & \cite{Ref79} \\
MCF 7 & Cancerous (Low) & 20--101 kHz & Resistance & 6.23--2.44 k$\Omega$ & \cite{Ref79} \\ \addlinespace

MCF 10 A & Healthy & 100 kHz & Specific membrane capacitance & 1.94 $\pm$ 0.14 $\mu$F/cm$^2$ & \cite{Ref80} \\
MCF 7 & Cancerous (Low) & 100 kHz & Specific membrane capacitance & 1.86 $\pm$ 0.11 $\mu$F/cm$^2$ & \cite{Ref80}\\
MDA-MB 231 & Cancerous (High) & 100 kHz & Specific membrane capacitance & 1.63 $\pm$ 0.17 $\mu$F/cm$^2$ & \cite{Ref80} \\
MDA-MB 435 & Cancerous (Met.) & 100 kHz & Specific membrane capacitance & 1.57 $\pm$ 0.12 $\mu$F/cm$^2$ & \cite{Ref80}\\ \addlinespace

AML 2 cells & Cancerous & 20--400 kHz & Specific membrane capacitance & 12 $\pm$ 1.44 mF/m$^2$ & \cite{Ref81} \\
HL-60 cells & Cancerous & 20--400 kHz & Specific membrane capacitance & 14.5 $\pm$ 1.75 mF/m$^2$ & \cite{Ref81} \\ \bottomrule
\end{tabularx}
\end{sidewaystable}

Similar studies by Roberts et al. (2005) investigated the progression of mouse ovarian surface epithelial (MOSE) cells \cite{Roberts2005}. This work analyzed cellular and molecular changes across the different stages, from pre-malignant, non-tumorigenic states to highly aggressive malignant phenotypes. The model includes four transitional states including MOSE-E (early preneoplastic stage) and MOSE-L (the most aggressive stage) with high tumorigenic potential. 

In addition, to dielectric spectroscopy, studies on dielectrophoresis (DEP) – a phenomenon in which a force is exerted on dielectric particle in a non-uniform electric field has also been employed to investigate the different stages of MOSE cells \cite{Trainito2019}. The results indicate that as malignancy progresses the membrane capacitance decreases from approximately 26 mF/m2 to 15 mF/m2. This decline highlights the difference in dielectric properties and membrane composition between malignant and non-tumorous cells, reinforcing the potential of electrical characterization methods in distinguishing cancer progression. 

While dielectric properties provide a promising biophysical basis for distinguishing between malignant and non-malignant cells, the complexity and variability of the measured parameters necessitate robust computational tools for accurate classification. Artificial Intelligence (AI) and Machine Learning (ML) algorithms have been increasingly adopted to analyze bioelectrical data. These models can learn patterns from features such as membrane capacitance, conductivity, and impedance spectra to reliably differentiate between healthy and cancerous cells. Among various ML techniques, models such as Random Forest (RF), Support Vector Machine (SVM) and K-Nearest Neighbor (KNN), and Neural Network have shown promising results in biomedical classification tasks due to their ability to handle non-linear data and high dimension feature space \cite{Breiman2001},\cite{Liu2012},\cite{Peterson2009},\cite{Kramer2013},\cite{Hearst1998},\cite{Noble2006},\cite{Hossain2021}.  


\subsection{Machine Learning Models in Bioelectrical Cancer Classification}\label{subsec2}
Late diagnosis of breast cancer can delay treatment and typically requires skilled radiologists to analyze diagnostic data. This process is time consuming and resource intensive. As an alternative, automated classification systems based on ML offer the potential for faster and more accurate preliminary diagnoses, particularly in settings with limited access to medical specialists. 

Supervised ML algorithms can be trained using datasets compiled from literature and clinical studies to capture diverse patient profiles. In this study, the input features are restricted to key electrical properties -dielectric constant, frequency, conductivity, and characteristic relaxation time constant- as the primary aim is to examine how these properties evolve during the transformation of normal cells to tumorous ones. Understanding these bioimpedance signatures may contribute to future strategies for electrically guided cancer detection or intervention \cite{Hossain2021},\cite{Hossain2024},\cite{Hossain2021b},\cite{Hossain2022},\cite{Hossain2021c}. To improve the model’s robustness, future work will explore the use of simulation tool to generate synthetic data and expand the dataset. 

In this work, we evaluate three supervised algorithms – Random Forest (RF), Support Vector Machine (SVM), K- Nearest Neighbor (KNN) – because they complement one another in handling non-linear high dimensional bioelectrical features. RF offers inherent bootstrapping and feature importance score and SVM is implemented with a radial-bias -function kernel which is effective with small datasets with decision boundaries. 

\subsection{Understanding Random Forest Algorithm}\label{subsec3}

Random Forest (RF) is a supervised ML algorithm that constructs an ensemble of decision trees (DT) to improve prediction accuracy and reduce overfitting \cite{Salman2024},\cite{Halabaku2024}. Each decision tree is trained on a random subset of data and selects a random subset of features to make its prediction. In classification tasks, the final output is determined by majority voting among the individual trees, enhancing robustness and generalization of the model.  This approach not only enhances prediction accuracy but reduces the risk of overfitting, which is critical when working with biological data that may vary across patients or experimental setups. In the context of this study, RF offers a powerful framework for classifying cells based on their dielectric signatures. Supporting the overarching goal of early, non-invasive cancer detection using bioimpedance characteristics. 
\subsection{Understanding Support Vector Machine (SVM)}\label{subsec3}

Another widely used classification model in the biomedical signal analysis is the Support Vector Machine (SVM) \cite{Pisner2020},\cite{Meyer2001}. SVM are particularly effective in high dimensional spaces and are known for their robustness in handling small-medium sized datasets with clear margins of separation. By constructing a hyperplane that best divides the data into classes, SVM can classify dielectric properties of tissues. When data is not linearly separable, often in the case with biological signals, SVM employs kernel functions (e.g. radial basis function, polynomial kernel) to project the data into a higher dimensional feature space where linear separation becomes feasible. 

SVM is also robust to overfitting, especially in high dimensional spaces, and performs well when there is a clear margin of separation between classes. For breast cancer detection, this means SVM can effectively distinguish between benign, normal and malignant cells by learning subtle differences in their electrical properties. Additionally, SVM can handle both binary and multiclass classification makes it suitable for distinguishing different stages or types of abnormal cellular behavior. 
Another advantage is that SVM models are relatively interpretable compared to more complex deep learning approaches, which is valuable in clinical contexts where understanding the decision boundary is important for trust and transparency. However, SVM can be computationally intensive for large datasets and require careful tuning of hyperparameters (e.g. kernel type, regularization parameter C, and gamma in RBF kernels) to achieve optimal performance. 

\subsection{Understanding K-Nearest Neighbor}\label{subsec3}

K-Nearest Neighbor (KNN) is a non-parametric instance-based supervised machine learning algorithm that classifies data points based on the classes of their nearest neighbors in the feature space \cite{Bezdek1986},\cite{Zhang2016}. The algorithm relies on a user-defined parameter, k, which determines the number of neighboring points considered when assigning a class label. KNN uses various distance metrics, such as Euclidean distance, Manhattan distance, and Minkowski distance-to compute the similarity between instances \cite{Iqbal2020}. These distance functions are critical in determining the neighborhood structure and can significantly impact classification performance. 
In the case of breast cancer detection using bioimpedance signatures, KNN can serve as a simple effective baseline model for distinguishing between normal, benign and malignant cells. The algorithm’s strength lies in its simplicity and its adaptability to both linear and nonlinear data distributions without requiring explicit model training. Moreover, it can capture local patterns and is particularly useful in cases where the decision boundary is irregular or data driven. 
For this study, different values of k will be tested to identify the optimal combination that yields the highest classification accuracy, minimizing false positives or negatives, and improve model reliability. Although KNN can be computationally expensive during inference, especially on large datasets, it remains a useful method for comparative analysis in non-invasive bioimpedance -based cancer classification tasks. 
\subsection{Physics Based Machine Learning Approach}
In the second part of the study, the original dataset is augmented using a physics-based modeling framework to derive additional dielectric descriptors from the measured parameters. Specifically, secondary features such as (i) loss tangent (ii) imaginary permittivity and (iii) charge relaxation time are computed from the primary dielectric properties, namely electrical conductivity and relative permittivity of the biological cell samples. The objective of incorporating these physics-derived parameters is to enrich the feature space and provide additional physical insight into dielectric relaxation and energy dissipation mechanisms within the cells, thereby improving the robustness of the machine learning model when limited primary descriptors are available. 

\section{Data Collection and Methodology}\label{sec2}

For this preliminary study, a sample dataset was compiled from 20 scholarly publications with 535 datasets using the keyword “dielectric properties of cancer cells” in academic search engines such as Google Scholar and PubMed Central (PMC). 
Each manuscript was screened for quantitative data on dielectric properties of normal (healthy), benign and malignant cells. The primary parameters extracted included relative permittivity ($\varepsilon_r$), characteristic relaxation time constant ($\tau_p$), and conductivity ($\varepsilon$). The frequency range of the datasets spanned from 150 kHz to 20 GHz. 
Initial analysis revealed that both relative permittivity and conductivity were significantly higher in breast cancer cells compared to their healthy counterparts \cite{Hossain2020},\cite{Hossain2021}. This difference highlights the potential of dielectric properties as discrimination biomarkers for cancer classification. 
These compiled dielectric parameters were then pre-processed for ML model development. Data from different studies were standardized to ensure consistent units and measurement scales. The resulting standardized datasets was subsequently divided into training and testing subsets, forming the foundation for evaluating the classification performance of various supervised learning algorithms, including RF, SVM and KNN.  
Three supervised ML algorithms were evaluated using the compiled datasets: RF, SVM and KNN. The performance of these algorithms was assessed using four key metrics: accuracy (ACC), F1 score, recall and precision \cite{Asif2021},\cite{Hossain2025},\cite{Hossain2025b},\cite{hossain2025c}. Accuracy measures the proportion of correct prediction relative to the total prediction made by the model and is generally more reliable for balanced datasets. However, given the limitations and potential imbalance in the literature -derived datasets, recall and precision provide additional insight. Recall is defined as the ratio of correctly classified actual positives to all actual positives, while precision is the ratio of correctly classified actual positives to all predicted positives \cite{Flach2015}. The F1-score, calculated as the harmonic means of precision and recall, offers a balanced evaluation of both metrics \cite{Diallo2024}. 
As outlined earlier, the primary objective of this study is to compare the predictive performance of the three models, to determine the pathological status of the breast cell, based on the permittivity and conductivity measurement. The algorithm will be used to optimize performance and the corresponding evaluation metrics will be used to identify the model with the highest predictive capability, thereby reducing the risk of false positives and false negatives. 
For the RF model, the scikit-learn library was used with default parameters \cite{Pedregosa2011}. The dataset was partitioned into 80\% training and 20\% testing subsets. The number of estimators varied from 1 to 500, while the maximum tree depth varied from 1 to 10. The corresponding F1 score, and accuracy were recorded, and the results are discussed in the following section. 
A similar procedure was applied to the SVM and the linear SVC models, also implemented using the scikit-learn library with default parameter (e.g. regularization parameter C=1). The maximum number of iterations varied between 10, 50, 100, 200, 500, 1000 and 2000. The corresponding evaluation metrics were analyzed to investigate the relationship between hyperparameter variation and performance improvement, particularly with respect to the F1 score. 

The KNN classifier was implemented using the scikit-learn library. The number of neighbors, k, varied across multiple magnitudes (e.g.1,3,5,7,9,11, and 15). For each value of k, the corresponding evaluation metrics, accuracy and F1 score, were computed on the test dataset. The results highlight the influence of the neighborhood size on classification performance, demonstrating the trade-off between model complexity and generalization ability. 

In the second stage of this study, physics-based model framework is implemented using the same datasets. The complete feature set is constructed by compiling six parameters: (i) measured frequency, (ii) electrical conductivity, (iii) relative permittivity, (iv) loss tangent, (v) imaginary permittivity, and (vi) relaxation time. A Random Forest classifier is then employed to evaluate the relative contribution of each parameter to the classification task. The model is trained using 500 estimators with a maximum tree depth of 15, and the corresponding feature importance weights are compared to determining the significance of each dielectric parameter in accurately predicting the cancer state.

Subsequently, a feature selection strategy is applied by imposing an importance threshold greater than 0.1 to retain only the most discriminative parameters. The reduced feature set is then used to retain the Random Forest model using the same hyper parameter configuration. The resulting model performance is evaluated using element wise classification metrics and validation testing curves to assess the improvement in prediction efficiency achieved through the incorporation of physics-informed feature selection. 

\section{Results and Discussion}\label{sec2}
 
The Random Forest (RF) classifier was evaluated by varying the number of estimators from 100 to 500 in increments of 100 as shown in Table \ref{tab1}. The corresponding performance metrics were computed for each configuration. Across the evaluated configurations, the precision recall, and overall classification accuracy converged to approximately 86-87\% indicating stable ensemble performance. 
To further investigate model complexity, the maximum tree depth varied from 1 to 10 while considering estimator values of 100 and 500. The highest classification accuracy ($\approx$87\%) and macro averaged F1 score were achieved at higher tree depths ($\ge$ 10), irrespective of the number of estimators. Increasing the maximum depth beyond 10 did not yield a statistically significant improvement in performance, indicating a saturation effect in ensemble learning capacity.
In contrast, shallow tree (maximum depth=1) resulted in reduced performance, with a F1-score of approximately 66\%, demonstrating underfitting due to limited model representation capability. Furthermore, no substantial performance improvement was observed for depths exceeding 10, suggesting that model complexity saturation occurs beyond this threshold. 
\begin{figure}[ht]
    \centering
    \includegraphics[width=1.1\textwidth]{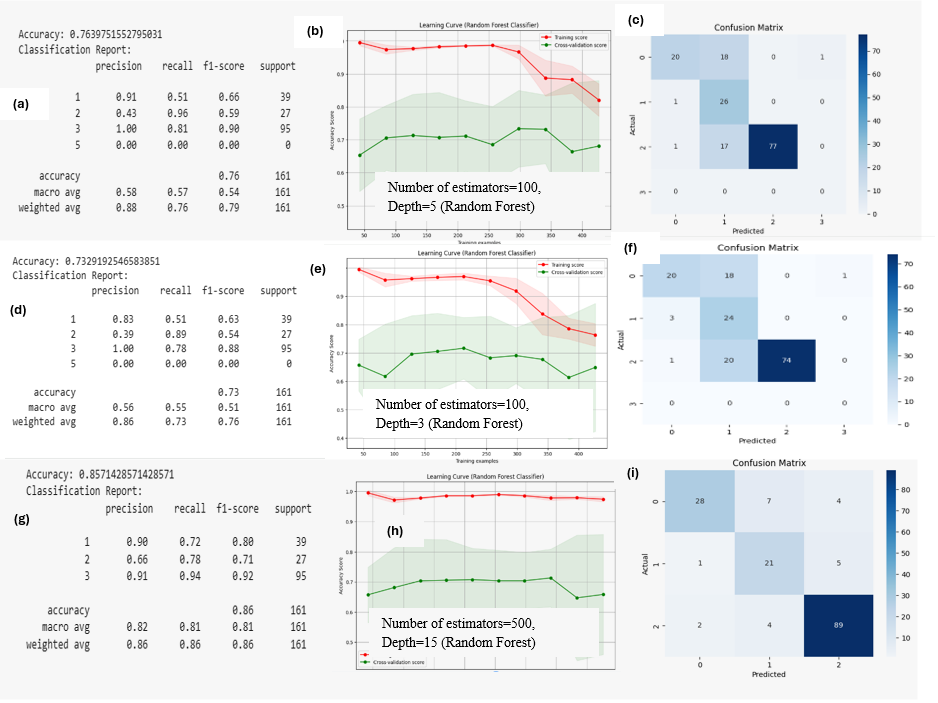} 
    \caption{\small{Comparative performance analysis of the Random Forest classifier for varying hyperparameter configurations. (a), (d), and (g) report class wise quantitative metrics; (b), (e), and (h) depict training versus cross-validation accuracy trends; and (c), (f), and (i) show confusion matrices. Case I : depth=5, estimators =100; Case II : depth=3, estimators =100; Case III: depth=15, estimators =500. }} 
    \label{fig:biomarkers1}
\end{figure}

Fig.~ \ref{fig:biomarkers1} presents the quantitative evaluation results, including (i) class-wise and macro-averaged performance metrics, (ii) the corresponding confusion matrices, and (iii) training and cross validation learning curves for three representative hyperparameter configurations:
Case I : Maximum depth=5, Number of estimators =100; Case II: Maximum depth=3, Number of estimators=100, and Case III : Maximum depth=15, number of estimators =500. 
In all three cases, the training accuracy consistently exceeds the cross-validation accuracy, indicating the presence of moderate overfitting. However, the performance gap decreases for deeper trees, suggesting improved generalization with increased model capacity. 
 A comparative assessment of Fig. 5 (a), (d) and (g) reveals that the RF classifier demonstrates strong discriminative capacity for malignant cells (Class III), achieving F1 scores exceeding 0.88 in Case III. Conversely, the classification performance for benign cells (Class II) remains comparatively lower, with F1 scores of approximately 0.59 and 0.54 for Case I and II, respectively and improving to approximately 0.71 in Case III.
The relatively lower benign -class sensitivity highlights the inherent difficulty in distinguishing benign from malignant cellular dielectric signatures. Enhancing benign-class discrimination is clinically significant as improved early stage differentiation may facilitate timely diagnosis and optimized therapeutic interventions. 

\begin{table}[h]
\caption{Classification Performance (F1 Score) for Various Tree Depths and Number of Estimators}\label{tab1}
\begin{tabular}{@{}llll@{}}
\toprule
Estimators & Depth = 1 & Depth = 5 & Depth = 10 \\
\midrule
100 & 66\% & 76\% & 86\% \\
200 & 66\% & 78\% & 86\% \\
300 & 70\% & 76\% & 87\% \\
400 & 69\% & 77\% & 87\% \\
500 & 69\% & 76\% & 86\% \\
\bottomrule
\end{tabular}
\end{table}
 
For linear SVC, the total computation time was slightly higher at 1.31 seconds. The evaluation metrics were significantly lower, with 53.85\%, accuracy, 28.99\% precision, and 37.69\% F1 score for 2000 iterations. This demonstrates that the linear SVM is not accurate predictive model. The best performance for the SVM model was obtained with a sigmoid kernel and the default regularization parameter (C=1), yielding 76.57\% precision, 69.23\% accuracy and 64.4\% F1 score. In contrast, the worst performance was observed for the RBF kernel at C=15, where accuracy fell to 36.8\%. Interestingly, for the polynomial kernel with C=10, the model achieved a high precision of 90\% but a poor F1 score of only 37.69\%, indicating strong bias towards certain classes. This highlights the trade-off between precision and balanced classification performance in certain hyperparameter settings.
 
The datasets were then evaluated using different numbers of neighbors (k), ranging from 1 to 20 and including 50, with training proportions of 70\% and 80\% respectively. The highest F1 score is approximately 78\% and was achieved at k=2. However, the 70-30\% training test split proved more reliable, as it reduced the risk of overfitting.  For k=5, the F1 score remained above 70\%, but performance declined for higher values of ‘k’.

\begin{figure}[ht]
    \centering
    \includegraphics[width=1.0\textwidth]{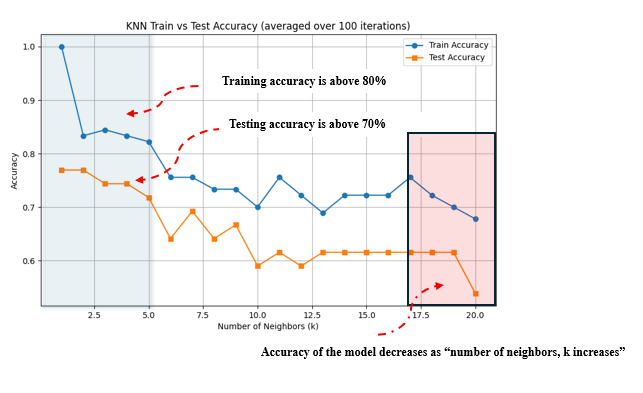} 
    \caption{\small{K-Nearest Neighbor (KNN) train vs. test accuracy model graph for 100 iterations showing the variation in evaluation metrics with number of k varied between 1 and 20.}} 
    \label{fig:knn_accuracy}
\end{figure}

Fig.~~ \ref{fig:knn_accuracy} compares kNN accuracy across different values of $k$ ($1$--$21$) over $100$ iterations using a $70$:$30$ train:test split. The model achieved higher accuracy (over $70\%$) for $k \leq 5$, but performance declined significantly for $k \geq 19$. A comparison of training versus testing accuracy further highlights the model’s behavior: (i) at $k=1$, the training accuracy reached $100\%$, while testing accuracy was only $76\%$, indicating strong overfitting; (ii) for $k \geq 10$, training accuracy dropped to about $70\%$, while testing accuracy fell below $60\%$, showing clear underfitting; (iii) the range $k=3$ to $5$ demonstrates better generalization, with training accuracy of approximately $82$--$85\%$, and testing accuracy of $72$--$75\%$ where the gap between training and testing accuracy was notably smaller.

\begin{figure}[ht]
    \centering
    \includegraphics[width=0.9\textwidth]{"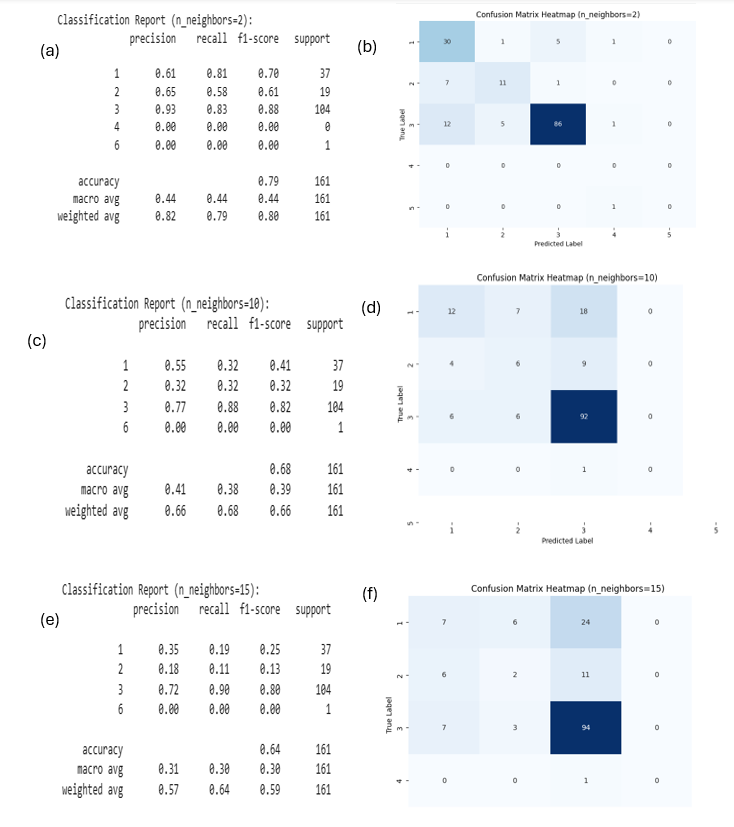"} 
    \caption{\small{Performance evaluation of the K-Nearest Neighbor (KNN) classifier for different neighborhood sizes. (a), (c) and (e) show the classification reports indicating precision, recall, and F1 score for each class, while (b), (d) and (f) present the corresponding confusion matrices. The models are evaluated for k=2, k=10, and k=15, respectively}} 
    \label{fig:knn_accuracy_1}
\end{figure}

Fig.~ \ref{fig:knn_accuracy_1} presents the performance evaluation of the K-Nearest Neighbor (KNN) classifier for different values of the neighborhood parameters, k. Subfigures (a), (c), and (e) illustrate the corresponding classification reports showing class-wise precision, recall, and F1 score, while subfigures (b), (d) and (f) present the associated confusion matrices. The classification is performed across all three cell categories: Class I (normal), Class II (benign), and Class III (malignant). 

The results indicate that the classifier performs comparatively well for the malignant cell class (Class III), achieving an F1 score of approximately 0.88 when k=2. This improved performance can be attributed to the relatively large number of samples available for this class in the dataset. However, as the number of neighbors increases to k=10 and k=15, the classification performance for the benign class (Class II) deteriorates significantly, as reflected by the reduction in precision and recall values. 

\begin{table}[ht]
\caption{Performance comparison of supervised learning models and their corresponding evaluation metrics}
\label{tab:model_comparison}
\centering
\begin{tabular}{@{}lll@{}}
\toprule
\textbf{Supervised Machine Learning Model} & \textbf{F1-score} & \textbf{Hyperparameters} \\ \midrule
Random Forest (RF) & 87\% & Maximum depth = 10, \\
 &  & Number of estimators = 400 \\ \addlinespace
Support Vector Machine (SVM) & 64.41\% & Regularization parameter, C = 1 \\
 &  & Sigmoid kernel \\ \addlinespace
K-Nearest Neighbor (KNN) & 79\% & $k = 2$, number of iterations = 100 \\ \bottomrule
\end{tabular}
\end{table}

Table~\ref{tab:model_comparison} presents the evaluation metrics for the supervised ML models. The RF achieved the highest accuracy of 90\% and F1 score =88.3\% with a maximum depth of 3 and 4 and 100 estimators. In contrast, the support vector machine (SVM) attains a lower accuracy of 66\% and F1 score of 64.4\%. Future work will focus on extending the model by incorporating additional parameters, such as capacitance as well as datasets derived from simulation models.

\begin{figure}[ht]
    \centering
    \includegraphics[width=1.0\textwidth]{"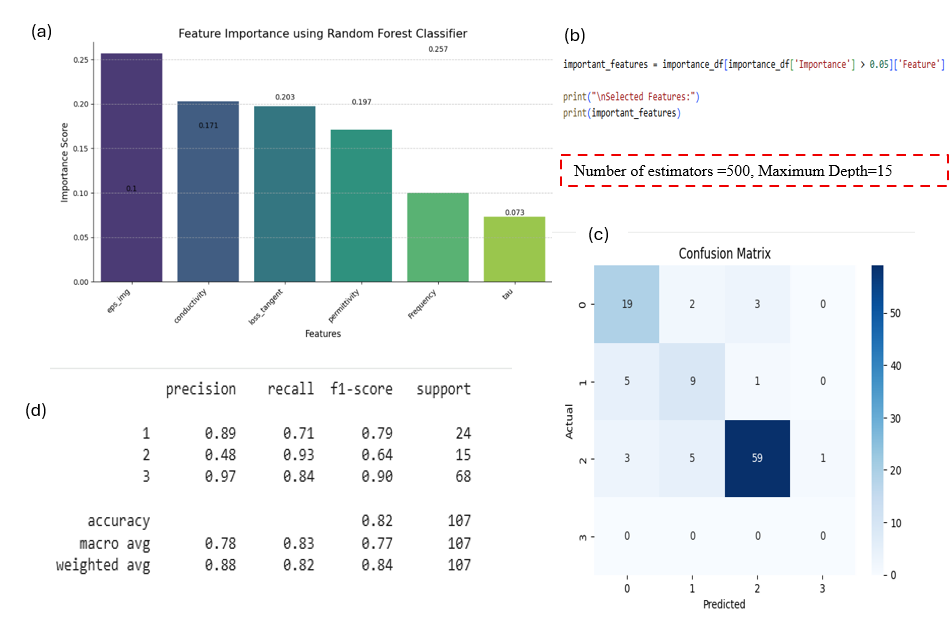"} 
    \caption{\small{(a)Feature importance scores obtained using Random Forest classifier for the extended dielectric feature set. (b) Feature selection algorithm used to retain parameters above the importance threshold (c) Confusion matrix obtained for the optimized Random Forest model with 500 estimators and maximum depth of 15. (d) Corresponding classification report showing precision, recall and F1 score for each cell class. }} 
    \label{fig:evaluation_results}
\end{figure}

Fig.~\ref{fig:evaluation_results} presents the evaluation results obtained after incorporating additional physics-derived dielectric parameters into the dataset. Fig.~\ref{fig:evaluation_results}(a) illustrates the relative feature importance obtained using the Random Forest Classifier. The height of each bar represents the contribution of the corresponding feature towards the classification task. From the feature importance analysis, it can be observed that the imaginary component of permittivity ($\varepsilon''$), electrical conductivity ($\sigma$), and loss tangent ($\tan \delta$) exhibit high importance scores, indicating that dielectric loss-related parameters play a significant role in distinguishing between different cell types.

Initially the classification model was trained using the complete set of features including: (i) measured frequency, (ii) conductivity, (iii) relative permittivity, (iv) loss tangent, (v) imaginary permittivity, and (vi) relaxation time ($\tau$). The feature selection process retained parameters with an importance score of greater than 0.005, as illustrated in the algorithm shown in Fig.~\ref{fig:evaluation_results} (b). The Random Forest Classifier was implemented using 500 estimates and a maximum depth of 15.

The corresponding classification performance is summarized in Fig.~\ref{fig:evaluation_results}(c) and (d) , which present the confusion matrix of element-wise evaluation metrics, respectively. The model achieved an overall classification accuracy of approximately 82\%. For malignant cell classification (Class III) the model obtained an F1 score of 90\%, precision of 97\% and recall of 84\%. Although the overall accuracy is slightly lower than the results obtained from the previous Random Forest model using the same hyperparameter, a notable improvement is observed in the recall score for the benign cell class, which increased from 75\% to 93\%. 
The recall metric is particularly important in biomedical classification problems, as it measures the ability of the machine learning model to correctly identify all relevant instances of a given class. An improvement in recall therefore indicates enhanced sensitivity of the model in detecting the corresponding cell state, which is critical for early diagnosis applications. 

\begin{figure}[ht]
    \centering
    \includegraphics[width=1.2\textwidth]{"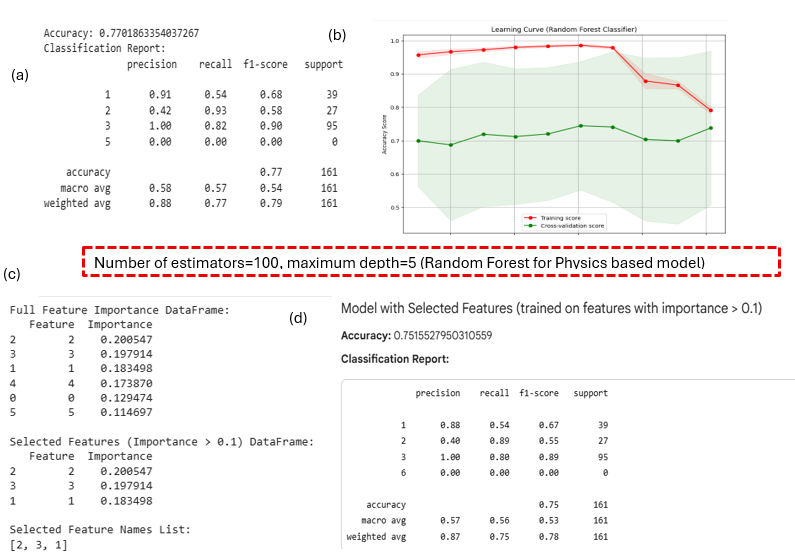"} 
    \caption{\small{Evaluation of the physics informed Random Forest model with reduced model complexity (estimators=100, maximum depth=5) : (a) Classification report for the full feature set; 
    (b) Learning curve showing training and validation scores;
    (c) Feature importance ranking of the dielectric parameters; and 
    (d) Classification performance using only the selected high-importance physics derived parameters. }} 
    \label{fig:results_1}
\end{figure}

To further evaluate the influence of physics-informed feature selection and model complexity on the classification performance, an additional analysis was performed using a reduced Random Forest architecture with the number of estimates set to 100 and maximum tree depth limited to 5 as shown in Fig. ~\ref{fig:results_1}. The corresponding feature importance weights for all dielectric parameters are summarized in Fig. ~\ref{fig:results_1}(c). Based on a feature importance threshold of 0.1, three dominant parameters were identified (i) imaginary permittivity (= 0.19), (ii) electrical conductivity (=0.20) and (iii) relative permittivity (=0.18). 

Fig. ~\ref{fig:results_1} (a) presents the classification performance obtained when the complete feature set was used in the Random Forest model. The model achieved an overall accuracy of approximately 77\%, which represents a marginal improvement compared to the earlier model accuracy of 76\%. However, analysis of the learning curves shown in Fig. ~\ref{fig:results_1}(b) indicates a reduction in overfitting behavior, as evidenced by the smaller gap between the training and validation scores. 
Subsequently, the model was retained using only the three dominant physics-informed parameters identified through the feature importance analysis. The resulting model achieved an overall accuracy of approximately 75\%, indicating only a slight reduction in predictive performance. Furthermore, no significant improvement was observed in the corresponding F1 scores for the individual classes. 

These results suggest that while physics-derived dielectric parameters contribute valuable interpretability and reduce model complexity, their inclusion does not necessarily lead to significant improvements in classification accuracy for the current dataset. 

\section{Conclusion and Future Work}\label{sec2}

Bioelectrical differences in cell properties such as relative permittivity, conductivity, and characteristic time constant as a function of frequency demonstrated significant variation between malignant and healthy cells. These distinctions highlight the potential of this technology for applications in cell classification and diagnosis. 
In this study, twenty scholarly articles were reviewed, and their datasets were compiled to provide a comprehensive overview and quantitative analysis of cell properties. Three supervised ML models- Random Forest (RF), Support Vector Machine (SVM), and K-Nearest Neighbor (KNN) model were evaluated for their effectiveness in predicting and classifying cell types.  Model hyperparameters including :(i) maximum depth and number of estimators for RF, (ii) kernel type and regularization parameters for SVM, and (iii) number of neighbors for KNN, were tuned to improve the predictive performance. Accuracy and F1 scores served as evaluation metrics to compare model effectiveness. 
The second part of the study investigated the potential of integrating physics-derived dielectric descriptors with machine learning algorithms for the classification of biological cell states. The results demonstrate that dielectric loss-related parameters such as imaginary permittivity, conductivity, and loss tangent exhibit strong discriminative capability for distinguishing between normal, benign, and malignant cells. Although the inclusion of physics-informed features improves model interpretability and reduces overfitting tendencies, the overall classification accuracy remains comparable to models trained solely on primary dielectric parameters. These findings highlight the importance of combining physics-based insights with machine learning approaches for informed understanding of dielectric spectroscopy- based biomedical diagnostics. 

Future research should focus on expanding the parameter space by integrating additional discriminative bioelectrical features and leveraging datasets derived from both experimental and simulation studies. Advanced hyperparameter optimization techniques, such as grid search and Bayesian methods, could further enhance predictive performance. Moreover, the development of hardware prototypes with embedded microelectrode arrays and real-time control systems may enable point-of-care applications for rapid and reliable diagnosis.







\section*{Declarations}

\begin{description}
    \item[Conflict of Interest
    ] Shadeeb Hossain is the sole proprietor of Shadeeb Engineering Lab, which may have a commercial interest in technologies related to this research.

    \item[Funding Declaration] No funding was received for this study.
    
    \item[Author Contributions] All authors contributed to the study conception and design. 
\end{description}






\bibliography{sn-bibliography}

\end{document}